\documentclass[cameraready]{Interspeech}

\title{Plug-and-Steer: Decoupling Separation and Selection \\
in Audio-Visual Target Speaker Extraction}

\author[affiliation={1}, equalcontribution]{Doyeop}{Kwak}
\author[affiliation={1}, equalcontribution]{Suyeon}{Lee}
\author[affiliation={1}]{Joon Son}{Chung}

\address{
    $^1$ Korea Advanced Institute of Science and Technology, South Korea
}

\email{dobbyk@kaist.ac.kr, syl4356@kaist.ac.kr, joonson@kaist.ac.kr}


\usepackage{comment}
\usepackage{booktabs}
\usepackage{multirow}
\usepackage{graphicx}
\usepackage{algorithmic}
\usepackage{algorithm}
\usepackage[table]{xcolor}
\usepackage{pifont}
\usepackage{url}
\usepackage{relsize}
\usepackage{hyperref}

\usepackage{fontawesome5}
\usepackage{makecell}
\usepackage{cite}

\begin{document}

\maketitle

\begin{abstract}
The goal of this paper is to provide a new perspective on audio-visual target speaker extraction (AV-TSE) by decoupling separation and target selection. Conventional AV-TSE systems typically integrate audio and visual features deeply to re-learn the entire separation process, which can act as a fidelity ceiling due to the noisy nature of in-the-wild audio-visual datasets. To address this, we propose Plug-and-Steer, which assigns high-fidelity separation to a frozen audio-only backbone and limits the role of the visual modality strictly to target selection. We introduce the Latent Steering Matrix (LSM), a minimalist linear transformation that re-routes latent features within the backbone to anchor the target speaker to a designated channel. Experiments across four representative architectures show that our method effectively preserves the acoustic priors of diverse backbones, achieving perceptual quality comparable to that of the original backbones. Audio samples are on the demo page.\footnote{\href{https://plugandsteer.github.io}{https://plugandsteer.github.io}}
\end{abstract}

\section{Introduction}
Speech separation is the task of isolating individual voices from a recording where multiple people are speaking simultaneously—a challenge often referred to as the ``Cocktail Party Problem''. In recent years, the performance of Audio-only Speech Separation (AOSS) models has reached an impressive turning point with advanced architectures~\cite{wang2023tf, zhao2024mossformer2, li2024spmamba, jiang2025dpmamba}. These systems can now separate overlapping speech with remarkable clarity, often rivaling or even exceeding theoretical limits of acoustic quality~\cite{lutati2022sepit}. However, despite these milestones, audio-only models face a fundamental hurdle in real-world deployment: permutation ambiguity. They remain blind to the target's identity, unable to automatically determine which output channel contains the voice the user intends to hear.

To resolve this ambiguity, conventional Audio-Visual Target Speaker Extraction (AV-TSE) architectures typically integrate audio and visual features deeply within the model via cross-attention~\cite{lee2024seeing,li2024iianet,lin2023avsepformer, sato2021multimodal}, concatenation~\cite{pan2022usev,zhao2025clearervoice,li2023rethinking}, or tailored architectures for deep fusion~\cite{gao2021visualvoice,li2024ctcnet,pegg2024rtfs,tao2025seanet}. This design paradigm assumes that effective extraction requires re-learning the entire separation process to leverage visual cues for both (1) identifying the target and (2) refining the separation quality through lip-sync information. While this joint optimization can be beneficial, we question whether using visual cues to further improve acoustic separation—a task AOSS models already perform with significant proficiency—always yields a net gain. Large-scale audio-visual datasets collected in the wild, such as LRS2~\cite{chung2017lip} and VoxCeleb2~\cite{chung2018voxceleb2}, often contain intrinsic noise and reverberation, providing sub-optimal supervision for high-fidelity extraction~\cite{chou2024av2wav}. In such cases, full-parameter training on noisy audio-visual data can act as a fidelity ceiling, potentially limiting the final output quality compared to what is achievable by pure AOSS models trained on studio-quality corpora.

In this work, we propose a shift in perspective: \textbf{Plug-and-Steer}. Instead of pursuing entangled multi-modal fusion to achieve separation and selection simultaneously, we decouple these roles to leverage the strengths of pre-trained acoustic engines. We assign high-fidelity separation entirely to a frozen AOSS backbone and limit the visual modality's role strictly to target selection. This approach is motivated by our observation that frozen AOSS models appear to possess a latent structure where output channels can be reordered at the feature level within a single separator block.
We demonstrate that this requires nothing more than a minimalist $C \times C$ linear transformation, termed a Latent Steering Matrix (LSM).

By treating the LSM as a ``steering wheel,'' we utilize a lightweight visual steering module to anchor the target speaker to a designated output channel. Unlike post-hoc selection—a straightforward strategy that identifies the target from fully decoded outputs—our mechanism is embedded directly within the feature flow. By reusing fine-grained latent features, Plug-and-Steer eliminates the redundant re-encoding or synchronization stages required by external selection pipelines. Furthermore, the LSM bridges the internal feature flow, allowing the routing logic to be optimized directly through signal-level reconstruction losses. This enables a direct gradient flow that ensures more stable target selection than detached classification. 

Ultimately, this work serves as a proof-of-concept demonstrating that well-established AOSS backbones can be transformed into a target extraction system at minimal cost. By offering a scalable blueprint for this adaptation, our framework allows extraction performance to scale naturally alongside the rapid evolution of underlying separation engines. Our primary contributions are summarized as follows:
\begin{itemize}
\item We analyze the latent structural properties of diverse AOSS architectures, showing that speaker identity is permutable via a simple linear transformation at the feature level.
\item We propose Plug-and-Steer, a framework decoupling separation from selection to preserve the high-fidelity acoustic priors of frozen AOSS models.
\item We demonstrate that our internal steering mechanism achieves comparable selection accuracy and better computational efficiency than post-hoc selection strategies.
\item 
Experiments across four representative architectures validate that our method transforms diverse AOSS backbones into AV-TSE systems while preserving their perceptual quality.

\end{itemize}
\section{Method}
\label{sec:method}
Our methodology decouples separation from selection by treating the frozen AOSS backbone as a high-fidelity engine steered by visual cues.
We first investigate whether a single linear transformation can re-route latent features within the backbone. We then leverage this mechanism to train a visual steering module that anchors the target speaker to a fixed output channel.

\subsection{Latent Steering Matrix (LSM): A steering wheel}
\label{subsec:def_swap}
We introduce a simple feature-level re-routing strategy to control output channel permutations of the AOSS backbone within the latent space.
Given an intermediate audio feature from the $i$-th separator block $f_i \in \mathbb{R}^{C \times T_a}$, we assume that speaker identity reordering can be approximated by a linear transformation $W \in \mathbb{R}^{C \times C}$.
We term $W$ the Latent Steering Matrix (LSM), and apply it 
as a residual transformation as follows:
\vspace{-1mm}
\begin{equation}
f_i' = (I + g \cdot W)f_i,
\label{eq:steering_op}
\end{equation}
where $I$ is the identity matrix and $g \in \{0, 1\}$ is a binary gate. 
When the gate is inactive ($g\!=0$), the features remain unchanged. Conversely, an active gate ($g\!=1$) triggers $W$ to induce a latent speaker swap, effectively permuting the output channels.

\begin{figure}
    \centering
    \includegraphics[width=\linewidth]{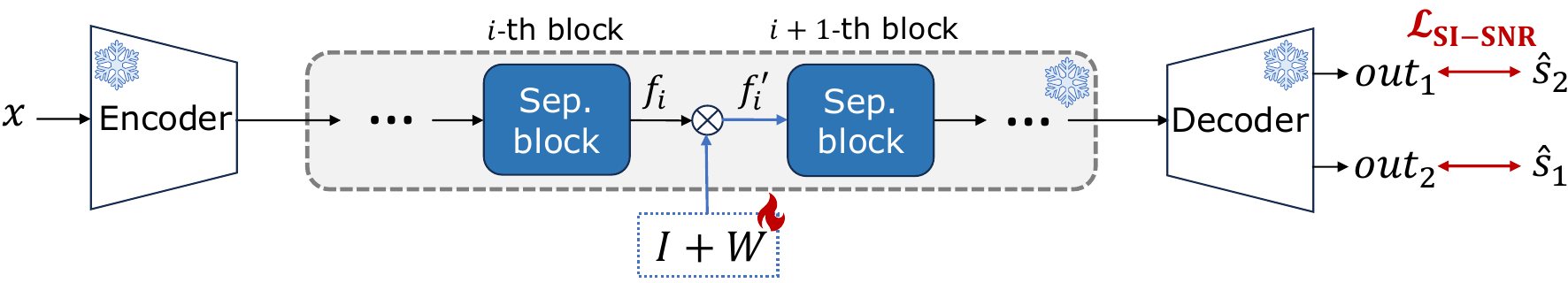}
    \caption{Training process of the latent steering matrix $W$ for the $i$-th separator block. $\hat{s}_1$ and $\hat{s}_2$ are the outputs from the first and second channels of the pre-trained audio-only backbone. }
    \label{fig:train_swap}
\end{figure}

\subsection{Training LSM}
\label{subsec:train_swap}
To learn the latent transformation required for speaker re-routing, we first train the LSM under a forced-swap condition by fixing $g=1$ within a frozen AOSS model.
Specifically, for a 2-speaker mixture where the pre-trained model outputs $(\hat{s}_1, \hat{s}_2)$, we train $W$ to manipulate the audio feature $f_i$ to produce the steered output: $(\hat{s}_2, \hat{s}_1)$. 
As shown in Fig.~\ref{fig:train_swap}, the channel-swapped predictions of the pre-trained backbone serve as the training targets.
The training objective is the sum of the negative Scale-Invariant Signal-to-Noise Ratio (SI-SNR)~\cite{le2019sdr} across both channels, computed between the steered outputs and the permuted reference signals.

\subsection{Visual steering module: Learning to steer}
\label{subsec:swap_tse}

\begin{figure}[!t]
    \centering
    \includegraphics[width=1\linewidth]{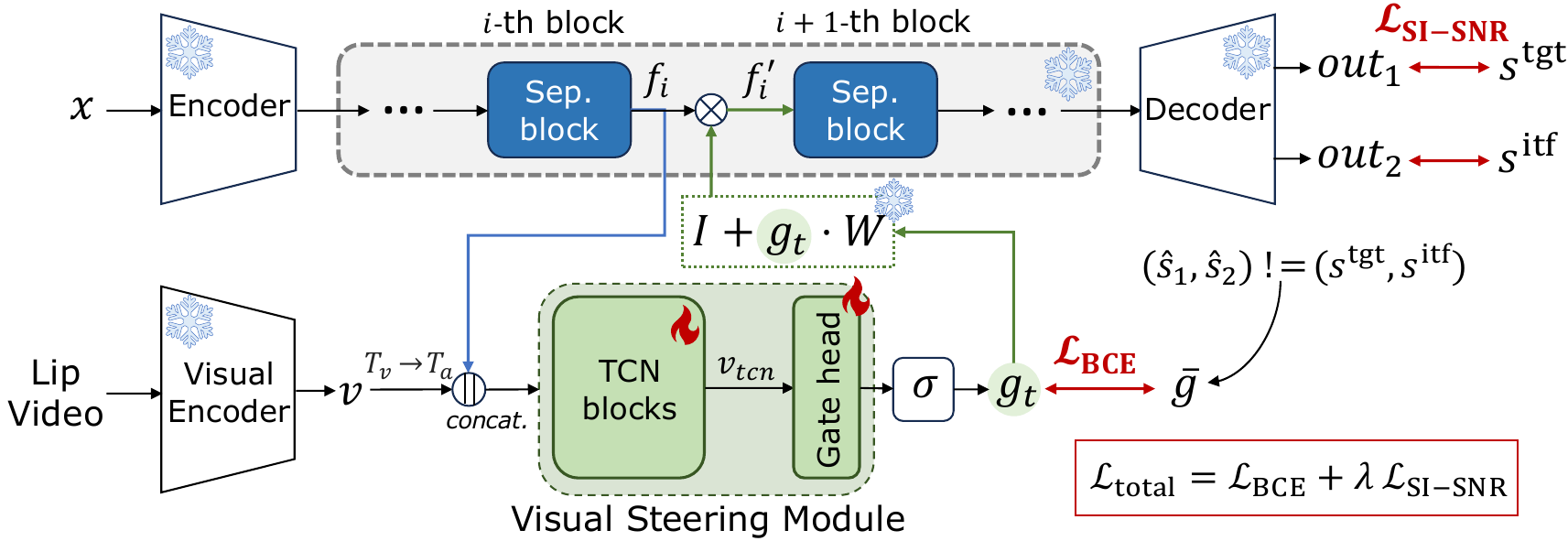}
    \caption{Training process of the AV-TSE module by learning the frame-wise gate value $g_t$. The visual steering module predicts the value $g_t$ to control the degree of steering.}
    \label{fig:train_av_swap}
    \vspace{-4mm}
\end{figure}

We adapt the AOSS model for AV-TSE by consistently routing the target speaker to a designated output channel, regardless of the backbone's initial output order. To this end, we learn a gate value to control the LSM based on visual cues.
As shown in Fig.~\ref{fig:train_av_swap}, we design a lightweight visual steering module that predicts a frame-wise gate value $g_t \in [0, 1]$ based on the latent audio features $f_i$ and the visual embedding $v \in \mathbb{R}^{T_v \times C_v}$ extracted from the target's lip motion. 

The temporal dimension of $v$ is linearly interpolated from $T_v$ to $T_a$ to match the resolution of the audio feature $f_i$.
We then concatenate the two features along the channel dimension. 
The joint feature is processed through a lightweight modified Temporal Convolutional Network (TCN) adapted from~\cite{luo2019conv}, which consists of two blocks, each containing three convolutional layers, followed by a sigmoid-activated gate head.
The output value $g_t$ dynamically modulates the steering operation based on Eq.~\ref{eq:steering_op}.
For backbones utilizing 2D latent spaces such as frequency-time grids, we first project the channel dimension to a reduced space $C_r$, and flatten the non-temporal dimensions into a 1D feature representation per time step.

We derive pseudo-labels $\bar{g}$ to supervise the gate by comparing the
frozen AOSS model's output permutations against
the desired target order.
Let the output of the AO backbone be $\hat{\mathbf{s}} = [\hat{s}_1, \hat{s}_2]^\top$, and 
the reference signals be $\mathbf{s}^{\text{ref}} = [s^{\text{tgt}}, s^{\text{itf}}]^\top$ for the target speech $s^{\text{tgt}}$ and interference $s^{\text{itf}}$. 
Defining $\mathbf{P}_g \in \mathbb{R}^{2 \times 2}$ as the permutation matrix indexed by $g \in \{0,1\}$ (identity for $g=0$, swap for $g=1$), the pseudo-label $\bar{g}$ is defined as:
\vspace{-2mm}
\begin{equation}
\label{eq:pseudo_label}
\bar{g} = \arg\max_{g \in \{0,1\}} \sum_{i=1}^{2} \mathrm{SI\text{-}SNR}\big([\mathbf{P}_g \,\hat{\mathbf{s}}]_i,\, s^{\text{ref}}_i\big),
\end{equation}

and used for the frame-wise binary cross-entropy loss $\mathcal{L}_{\text{BCE}}$.
The steered output $\hat{\mathbf{s}}^{\text{LSM}}$ is obtained by applying LSM with predicted $g_t$.
The SI-SNR loss $\mathcal{L}_{\text{SI-SNR}}$ is defined as the negative total SI-SNR between $\hat{\mathbf{s}}^{\text{LSM}}$ and the reference $\mathbf{s}$. 
The overall training objective for the visual steering module is given by:
\begin{equation}
\label{eq:total_loss}
\mathcal{L}_{\text{total}} = \mathcal{L}_{\text{BCE}}(g_t, \bar{g}) + \lambda \,\mathcal{L}_{\text{SI-SNR}}(\hat{\mathbf{s}}^{\text{LSM}}, \mathbf{s}).
\end{equation}
During this phase, only the visual steering module is updated, while the pre-trained backbone and the LSM remain frozen.

\section{Experimental Setup}
\subsection{Datasets}

We conduct our AV-TSE experiments on LRS2-2mix~\cite{zhao2025clearervoice}, a two-speaker speech separation benchmark partitioned into 20k ($\sim$\qty{23}{\hour}), 5k, and 3k samples for training, validation, and testing, respectively. Mixtures are synthesized by mixing pairs of utterances from different speakers in LRS2~\cite{chung2017lip} with random SNRs in [-5, 5] dB.
To analyze the effect of acoustic priors, we compare two pre-training configurations: a clean setup using the studio-quality Libri2Mix train-100 subset~\cite{cosentino2020librimix} ($\sim$ \qty{58}{\hour}) and an in-the-wild, noisier setup using the LRS2-2mix train set. 
Audio signals are sampled at \qty{16}{\kHz} in mono format and randomly truncated to \qty{3}{\second} during training with variance normalization. 
Visual inputs are 25 FPS grayscale sequences, obtained by center-cropping the original $224 \times 224$ LRS2 frames to $112 \times 112$.

\subsection{Training and evaluation details}
\noindent \textbf{Baseline models.}
We conduct our analysis across four representative speech separation models: Conv-TasNet~\cite{luo2019conv}, DPRNN~\cite{luo2020dprnn}, TF-GridNet~\cite{wang2023tf}, and MossFormer2~\cite{zhao2024mossformer2}. 
To compare our approach against the established AV-TSE baselines, we use AV-ConvTasNet~\cite{wu2019time}, AV-DPRNN~\cite{pan2022usev}, AV-TFGridNet~\cite{pan2023scenario}, and AV-MossFormer2~\cite{zhao2025clearervoice} with open-sourced weights pre-trained on LRS2-2mix\footnote{\url{https://github.com/modelscope/ClearerVoice-Studio/tree/main/train/target_speaker_extraction}} for evaluation. Our visual encoder follows the same architecture as the AV-TSE baselines and is initialized with lip-reading pre-trained weights.

\noindent \textbf{Metrics \& evaluation.}
We use the Scale-Invariant Signal-to-Distortion Ratio improvement (SI-SDRi)~\cite{le2019sdr} as a standard metric in speech separation, and DNSMOS~\cite{reddy2021dnsmos} and NISQA~\cite{mittag2021nisqa} as non-intrusive metrics to assess the perceptual quality of audio. For the audio-only backbones, we follow the Permutation Invariant Training (PIT) protocol~\cite{yu2017permutation}, selecting the output-target assignment that maximizes the SI-SDR.

\noindent \textbf{Inference.}
We apply a threshold $\tau=0.5$ to the averaged gate value $g$ during inference. 
For AV-TSE methods, we sequentially extract each speaker’s speech with the corresponding video.

\noindent \textbf{Training configurations.}
AOSS models are pre-trained using Adam~\cite{kingma2015adam} with an initial learning rate of $5\times 10^{-4}$. 
The learning rate is halved every 5 plateau epochs, with training terminating below $10^{-6}$.
The total batch size is 4, and the training objective is $\mathcal{L}_{\text{SI-SNR}}$.
After pre-training audio backbones, we optimize our Plug-and-Steer framework with a cosine annealing learning rate scheduler. With the AOSS backbone frozen, training is conducted in two stages: (1) LSM is trained for 10k steps without a warmup period; and (2) the visual steering module is trained for 100k steps, including a 1k-step warmup. 
$C_r$ for DPRNN and TF-GridNet is set to $16$, and $\lambda$ in Eq.~\ref{eq:total_loss} is set to $0.1$. All experiments, including inference, were conducted on NVIDIA RTX 4090 GPUs with an AMD EPYC 7543 CPU.

\subsection{Alternative adaptation strategies for AV-TSE}
\label{subsec:avtse_adaptation}
To evaluate the proposed steering mechanism, we compare it against two alternative strategies that adapt a pre-trained AOSS backbone for target extraction via residual connections.
These variants replace the gate head with a visual adapter that transforms $v_{tcn}$ into a residual feature: $f'_i = f_i + \text{Adapter}(v_{tcn})$. Unlike our approach, which simply re-routes existing representations, this method refines the acoustic feature itself. We consider both full fine-tuning, where the entire backbone is unfrozen, and partial adaptation, where the backbone remains frozen. For full fine-tuning specifically, the backbone learning rate is set to $5 \times 10^{-5}$ to mitigate catastrophic forgetting, while the adapter follows the standard training protocol.

\section{Results and Analysis}
\subsection{Layer-wise performance preservation}
\begin{figure}
    \centering
    \includegraphics[width=1\linewidth]{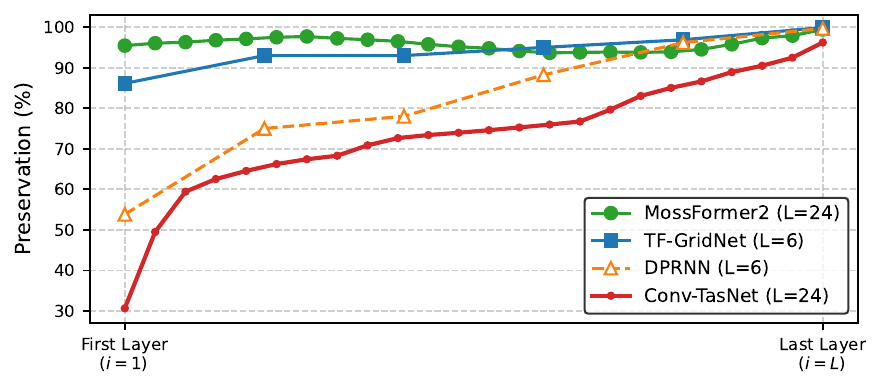}
    \vspace{-8mm}
    \caption{Layer-wise performance preservation rate (\%) across different AOSS backbones.  
    }
    \label{fig:layerwise_swap_perf}
    \vspace{-4mm}
\end{figure}
To identify the optimal location for latent manipulation, we evaluate the performance preservation rate across all separator blocks $i \in [1, \dots, L]$. This rate is defined as the ratio of the SI-SDRi achieved after applying the LSM relative to the original AO performance. As illustrated in Fig.~\ref{fig:layerwise_swap_perf}, the capacity for near-lossless swapping increases significantly with model depth. Applying the LSM at the final separator block yields the highest preservation rates across all architectures: 96.22\% for Conv-TasNet, 99.67\% for DPRNN, 99.91\% for TF-GridNet, and 99.43\% for MossFormer2. Notably, modern architectures such as TF-GridNet and MossFormer2 demonstrate a remarkable ability to maintain disentangled speaker features throughout the network, retaining 86.13\% and 95.44\% of their original performance even at their first separator blocks. These observations suggest that in advanced separation backbones, speaker identity is already distinctly encoded and manipulable from the early stages of processing, well before reaching the final output layers. Consequently, we utilize the final layer of each backbone for all subsequent AV-TSE experiments to ensure maximum fidelity.

\subsection{AV-TSE performance}
\subsubsection{Analysis of acoustic prior preservation}
\begin{table}[!t]
    \centering
    \resizebox{\columnwidth}{!}{
    \setlength{\aboverulesep}{1pt}
    \setlength{\belowrulesep}{2pt}
    \begin{tabular}{lc|cccc}
        \toprule
        {\textbf{Method}} & {\textit{\textbf{AO}} \faSnowflake } & \textbf{\# params.} & \textbf{SI-SDRi (dB)} & \textbf{DNSMOS} & \textbf{NISQA} \\ 
        \midrule
        \midrule
        \multicolumn{2}{l|}{\textit{Libri2Mix GT}} & - & - & 3.16 & 3.93 \\
        \multicolumn{2}{l|}{\textit{LRS2-2mix GT}} & - & - & 2.38 & 3.19 \\
        \midrule
        \midrule
        \multicolumn{6}{c}{{\textit{AO-oriented}}}\\
        \midrule
        \rowcolor{lightgray!40}
        \multicolumn{2}{l|}{Conv-TasNet~\cite{luo2019conv}} & {5.1M}& {7.12} & {2.35} & {2.31} \\
         + Residual & \ding{55} & 6.6M & \underline{11.72} & \underline{2.31} & \underline{2.57} \\
         + Residual & \ding{51} & 1.5M & 9.37 & 2.13 & 2.27 \\
         + \textbf{LSM (Ours)} & \ding{51} & 1.5M & 6.82 & 2.28 & 2.22 \\
        \rowcolor{lightgray!40}
        \multicolumn{2}{l|}{{DPRNN}~\cite{luo2020dprnn}} & {2.6M} & {7.77} & {2.35} & {2.37} \\
          + Residual  &  \ding{55}& 5.2M & \underline{11.66} & 2.28 & \underline{2.43} \\
          + Residual  & \ding{51}& 2.5M & 7.90 & 2.04 & 1.71 \\
         +  \textbf{LSM (Ours)}& \ding{51}& 2.0M & 7.74 & \underline{2.35} & 2.36 \\
        \rowcolor{lightgray!40}
        \multicolumn{2}{l|}{TF-GridNet~\cite{wang2023tf}} & {14.4M} & {14.81} & {2.80} & {4.32} \\
          + Residual  & \ding{55}& 17.0M & 13.38 & 2.36 & 3.17 \\
          + Residual  & \ding{51}& 2.6M & 13.09 & 2.56 & 3.20 \\
         + \textbf{LSM (Ours)}& \ding{51}  & 2.0M & \underline{14.79} & \underline{\textbf{2.79}} & \underline{\textbf{4.29}} \\
        \rowcolor{lightgray!40}
        \multicolumn{2}{l|}{MossFormer2~\cite{zhao2024mossformer2}} & {55.7M} & {12.68} & {2.79} & {3.47} \\
        + Residual  & \ding{55} & 57.3M & \underline{\textbf{15.54}} & 2.51 & 3.02 \\
        + Residual  & \ding{51} & 1.6M & 14.02 & 2.44 & 2.95 \\
        + \textbf{LSM (Ours)} &  \ding{51}& 1.6M & 12.65 & \underline{\textbf{2.79}} & \underline{3.47} \\
        \midrule
        \midrule
        \multicolumn{6}{c}{{\textit{AV-TSE Baselines}}}\\
        \midrule
        \multicolumn{2}{l|}{AV-ConvTasNet~\cite{wu2019time}} & 10.3M & 11.51 & 2.44 & 2.49 \\
        \multicolumn{2}{l|}{AV-DPRNN~\cite{pan2022usev}} & 4.1M & 11.97 & 2.40 & 2.58 \\
        \multicolumn{2}{l|}{AV-TFGridNet~\cite{pan2023scenario}} & 9.6M & 15.10 & 2.51 & \underline{3.53} \\ 
        \multicolumn{2}{l|}{AV-MossFormer2~\cite{zhao2025clearervoice}} & 57.3M & \underline{15.52} & \underline{2.64} & 3.06 \\ 
        \bottomrule
    \end{tabular}}
    \caption{{Performance of AV-TSE methods on LRS2-2mix grouped by audio-only (AO) backbone, where the AO backbones are pre-trained on Libri2Mix. \textit{\textbf{AO}} \faSnowflake \, indicates whether the pre-trained AO backbone is frozen or not. \textbf{\# params.} denotes the number of trainable parameters. Bold and underlined values represent global and backbone-specific best scores.}
    }
    \label{tab:libri_tse_results}
    \vspace{-8mm}
\end{table}
We first conduct a controlled analysis to investigate how Plug-and-Steer preserves acoustic priors compared to conventional adaptation strategies, specifically full and partial residual fine-tuning. As shown in Tab.~\ref{tab:libri_tse_results}, while residual-based methods may improve intrusive metrics such as SI-SDRi by aligning features with the LRS2-2mix distribution, they often suffer from degraded perceptual quality (DNSMOS and NISQA). This suggests that such methods compromise high-fidelity priors to accommodate the noisier ground truth of audio-visual datasets. In contrast, our LSM approach achieves target selectivity while largely retaining the original AO characteristics. In the case of TF-GridNet, residual adaptation results in suboptimal performance, even falling below the original AO backbone, which likely stems from architectural mismatches between the 2D feature space and the TCN-based steering module. These results indicate that conventional adaptation may require sophisticated, model-specific designs to bridge such structural gaps.  Conversely, since the objective of our approach is to simply steer existing latent features toward the designated output channel, it eliminates the need for complex feature manipulation and provides a robust, architecture-agnostic solution.

\subsubsection{Scaling with powerful AO backbones}
\begin{table}[!t]
    \centering
    \resizebox{0.87\columnwidth}{!}{
    \setlength{\aboverulesep}{1pt}
    \setlength{\belowrulesep}{2pt}
    \begin{tabular}{lc|ccc}
        \toprule
        {\textbf{Method}} & {\textit{\textbf{AO}} \faSnowflake} & \textbf{SI-SDRi (dB)} & \textbf{DNSMOS} & \textbf{NISQA} \\ 
        \midrule
        \midrule
        \rowcolor{lightgray!40}
        \multicolumn{2}{l|}{MossFormer2~\cite{zhao2024mossformer2}} & 15.42 & 2.88 & 3.88 \\
        + Residual & \ding{55} & \underline{\textbf{17.11}} & 2.53 & 3.28 \\
        + Residual & \ding{51} & 16.24 & 2.51 & 3.21 \\
        + \textbf{LSM (Ours)} & \ding{51}  & 15.40 & \underline{\textbf{2.88}} & \underline{\textbf{3.87}} \\
        \midrule
        \multicolumn{2}{l|}
        {AV-MossFormer2~\cite{zhao2025clearervoice}} & 15.52 & 2.64 & 3.06 \\
        \bottomrule
    \end{tabular}}
    \caption{AV-TSE performance on LRS2-2mix using a powerful backbone. 
    The baseline (AV-MossFormer2) is trained solely on LRS2-2mix.
    }
    \label{tab:strong_ao_tse_results}
    \vspace{-4mm}
\end{table}

The true potential of Plug-and-Steer is further evidenced when integrated with a more powerful backbone. We employ a MossFormer2 model pre-trained on a 107-hour high-fidelity corpus consisting of VCTK~\cite{yamagishi2019vctk}, LibriTTS~\cite{zen2019libritts}, and internal TTS data\footnote{\url{https://huggingface.co/alibabasglab/MossFormer2_SS_16K}}. As shown in Tab.~\ref{tab:strong_ao_tse_results}, applying our framework to this high-performance engine yields SI-SDRi levels comparable to established AV-TSE baselines while maintaining significantly higher perceptual fidelity. Although conventional fine-tuning can achieve higher SI-SDRi, it causes perceptual quality to decline to levels typical of models trained solely on noisy audio-visual data. By protecting acoustic latents from noisy supervision, Plug-and-Steer shows that the extraction quality of the system scales directly with the strength of the AO backbone, achieving high-fidelity extraction that conventional adaptation struggles to maintain.

\subsubsection{In-domain adaptation}
\begin{table}[!t]
    \centering
    \resizebox{0.85\columnwidth}{!}{%
    \setlength{\aboverulesep}{1pt}
    \setlength{\belowrulesep}{2pt}
    \begin{tabular}{lc|ccccc}
        \toprule
        {\textbf{Method}} & \textit{\textbf{AO}} \faSnowflake & \textbf{SI-SDRi (dB)} & \textbf{DNSMOS} & \textbf{NISQA} \\ 
        \midrule
        \midrule\rowcolor{lightgray!40}
        \multicolumn{2}{l|}{Conv-TasNet~\cite{luo2019conv}} & 11.50 & 2.30 & 2.32 \\
                                   + Residual & \ding{55} & \underline{11.33} & \underline{2.29} & \underline{2.33} \\
                                   + Residual & \ding{51}  & 11.28 & 2.28 & 2.31  \\
                                   + \textbf{LSM (Ours)} & \ding{51}  & 11.21 & 2.26 & 2.25  \\
        \rowcolor{lightgray!40}
        \multicolumn{2}{l|}{DPRNN~\cite{luo2020dprnn}}      & 12.60 & 2.38 & 2.55 \\
                                   + Residual & \ding{55} & \underline{12.59} & 2.34 & 2.51 \\
                                   + Residual & \ding{51} & 11.64 & 2.24 & 1.97 \\
                                   + \textbf{LSM (Ours)} & \ding{51} & 12.54 & \underline{2.36} & \underline{2.53} \\
        \rowcolor{lightgray!40}
        \multicolumn{2}{l|}{TF-GridNet~\cite{wang2023tf}} & 15.90 & 2.52 & 3.61 \\
                                   + Residual & \ding{55}  & 14.01 & 2.45 & 3.20 \\
                                   + Residual & \ding{51}  & 14.93 & 2.47 & 3.33 \\
                                   + \textbf{LSM (Ours)} & \ding{51}  & \underline{\textbf{15.87}} & \underline{\textbf{2.52}} & \underline{\textbf{3.60}}\\
        \rowcolor{lightgray!40}
        \multicolumn{2}{l|}{MossFormer2~\cite{zhao2024mossformer2}}& 15.07 & 2.52 & 3.15  \\
                                   + Residual & \ding{55} & \underline{15.54} & 2.51 & 3.02 \\
                                   + Residual & \ding{51}  & 15.03 & \underline{\textbf{2.52}} & 3.13 \\
                                   + \textbf{LSM (Ours)} & \ding{51}  & 15.02 & \underline{\textbf{2.52}} & \underline{3.14}\\
        
        \bottomrule
    \end{tabular}}
    \caption{In-domain AV-TSE performance on LRS2-2mix. Both AO and AV models are trained on the same data.
    } 
    \label{tab:main_tse_results}
    \vspace{-6mm}
\end{table}
Tab.~\ref{tab:main_tse_results} summarizes the in-domain adaptation performance on LRS2-2mix, where AO backbones are pre-trained and adapted within the same data distribution. In this scenario, the performance of the proposed LSM remains closely anchored to the original AO results, implying that the separation quality of the backbone acts as both a performance floor and an upper bound. This characteristic ensures that the inherent separation capabilities are effectively preserved, with the final extraction fidelity naturally defined by the pre-trained AO engine. Unlike conventional residual-based fine-tuning, which can be sensitive to specific architectures and often necessitates specialized tuning to avoid degradation, our approach maintains consistent stability across all backbones with minimal parameter updates. Although LSM may not always yield the absolute peak performance, its structural simplicity and minimal training requirements position it as a robust and practical baseline for transforming any pre-trained AO engine into a stable TSE system.

\subsection{Routing strategy and optimization}
\begin{table}[!t]
\centering
\resizebox{\columnwidth}{!}{
    \begin{tabular}{ll|cccr}
    \toprule
    \multirow{2}{*}{\textbf{Routing}} 
    & \multirow{2}{*}{\textbf{Method}} 
    & \multicolumn{2}{c}{\textbf{Accuracy (\%)}} 
    & \multirow{2}{*}{\textbf{FLOPs}} 
    & \multirow{2}{*}{\textbf{RTF}} \\
    
    \cmidrule(lr){3-4}
    & & \textbf{1 epoch} & \textbf{Best epoch} & & \\
    \midrule
    \midrule
    
    \multirow{3}{*}{Internal}
    & $\mathcal{L}_{\text{BCE}}$                 & 50.00 & 99.73 & \multirow{3}{*}{\textbf{256.82 G}} & \multirow{3}{*}{\textbf{0.147}} \\
    & $\mathcal{L}_{\text{SI-SNR}}$ + LSM        & 97.03 & 99.60 &  &  \\
    & $\mathcal{L}_{\text{BCE}}$ + $\mathcal{L}_{\text{SI-SNR}}$ + LSM  & \underline{99.01} & \underline{\textbf{99.93}} &  &  \\
    
    \midrule
    \multirow{2}{*}{Post-hoc}
    & LSE-C~\cite{Chung16a,prajwal2020lip} & \multicolumn{2}{c}{99.80} & \multirow{2}{*}{308.56 G} & \multirow{2}{*}{0.209} \\
    & LSE-D~\cite{Chung16a,prajwal2020lip} & \multicolumn{2}{c}{\underline{99.83}} &  &  \\
    \bottomrule
    \end{tabular}
    }
\caption{Internal routing vs. post-hoc selection. `Internal' denotes our routing module with various loss combinations; `Post-hoc' refers to SyncNet-score-based selection.}
\label{tab:routing_ablation}
\vspace{-8mm}
\end{table}
To validate our internal routing mechanism, we compare it against a post-hoc selection pipeline, which represents one of the most straightforward strategies for selecting a target speaker without retraining the AOSS backbone. In this cascaded setup, separator outputs are decoded and subsequently evaluated by an off-the-shelf SyncNet-based model~\cite{Chung16a}. 
The target is determined based on lip-sync consistency between the decoded output and the video, as measured by LSE-C and LSE-D scores~\cite{prajwal2020lip}. 
As shown in Tab.~\ref{tab:routing_ablation}, while our method achieves slightly higher routing accuracy, its primary advantage lies in system efficiency. Unlike post-hoc approaches that necessitate additional decoding, re-encoding, and potential preprocessing, Plug-and-Steer reuses fine-grained latent features directly from the backbone for routing. This elimination of redundant stages reduces the computational burden from 308.56 to 256.82 GFLOPs and improves the real-time factor from 0.209 to 0.147.

Furthermore, ablation results indicate that joint optimization with classification and reconstruction losses is essential for stable and accurate routing. While $\mathcal{L}_{\text{BCE}}$ alone suffers from unstable initial convergence and $\mathcal{L}_{\text{SI-SNR}}$ yields a slightly lower final accuracy, combining both objectives ensures both rapid convergence and the highest final accuracy. 
Crucially, the LSM serves as a functional bridge within the latent feature flow, enabling the use of reconstruction-based objectives for the routing module. Without this structural link, gradients from the final output could not propagate back to the steering logic, reducing target selection to a detached classification task. By facilitating this direct gradient flow, the LSM ensures that visual steering is optimized specifically for signal-level reconstruction, resulting in faster convergence and superior extraction performance.
\section{Conclusion}
In this work, we introduced Plug-and-Steer, a framework that decouples acoustic separation from target selection by redefining the visual modality as a selective router. By using a frozen audio-only backbone and a latent steering matrix, we achieve target selection while preserving the high-fidelity acoustic priors of the original engine. Our results demonstrate that while traditional fine-tuning suffers from noisy audio-visual ground truths, our steering approach effectively shields signal integrity. We expect that this decoupled strategy will provide a robust and scalable blueprint for high-fidelity extraction that adapts seamlessly to evolving speech separation engines.
\section{Acknowledgements}
This work was supported by the Institute of Information \& Communications Technology Planning \& Evaluation (IITP) grants funded by the Korean government (RS-2022-II220989, Development of Artificial Intelligence Technology for Multi-speaker Dialog Modeling).
\section{Use of Generative AI Disclosure}
During manuscript preparation, the authors used OpenAI's ChatGPT and Google's Gemini for language editing and stylistic refinement. These tools were not involved in conceptualizing the methods, designing experimental protocols, or interpreting scientific results. The authors reviewed and verified all suggestions and maintain full responsibility for the integrity and accuracy of the final content.

\bibliographystyle{IEEEtran}
\bibliography{mybib}

@string{accv =  "Proc. ACCV"}

@string{acmmm = "Proc. ACM MM"}

@string{asru = "IEEE Automatic Speech Recognition and Understanding workshop"}

@string{cvpr   = "Proc. CVPR"}

@string{icassp="Proc. ICASSP"}

@string{iclr = "Proc. ICLR"}

@string{icml =  "Proc. ICML"}

@string{is =  "Proc. Interspeech"}

@string{taslp =  "IEEE/ACM Trans. on Audio, Speech, and Language Processing"}

@string{tpami =  "IEEE Trans. on Pattern Analysis and Machine Intelligence"}

@inproceedings{lutati2022sepit,
  title={{SepIt}: Approaching a Single Channel Speech Separation Bound},
  author={Lutati, Shahar and Nachmani, Eliya and Wolf, Lior},
  booktitle=is,
  year={2022},
}

@inproceedings{zhao2025clearervoice,
  title={{ClearerVoice-Studio}: Bridging advanced speech processing research and practical deployment},
  author={Zhao, Shengkui and Pan, Zexu and Ma, Bin},
  booktitle=is,
  year={2025}
}

@inproceedings{li2024iianet,
  title={{IIANet}: An intra-and inter-modality attention network for audio-visual speech separation},
  author={Li, Kai and Yang, Runxuan and Sun, Fuchun and Hu, Xiaolin},
  booktitle=icml,
  year={2024}
}

@inproceedings{pegg2024rtfs,
  title={{RTFS-Net}: Recurrent Time-Frequency Modelling for Efficient Audio-Visual Speech Separation},
  author={Pegg, Samuel and Li, Kai and Hu, Xiaolin},
  booktitle=iclr,
  year={2024}
}

@article{tao2025seanet,
  title={Audio-Visual Target Speaker Extraction with Reverse Selective Auditory Attention},
  author={Tao, Ruijie and Qian, Xinyuan and Jiang, Yidi and Li, Junjie and Wang, Jiadong and Li, Haizhou},
  journal=taslp,
  year={2025}
}

@inproceedings{lee2024seeing,
  title={Seeing through the conversation: Audio-visual speech separation based on diffusion model},
  author={Lee, Suyeon and Jung, Chaeyoung and Jang, Youngjoon and Kim, Jaehun and Chung, Joon Son},
  booktitle=icassp,
  year={2024},
}

@article{pan2022usev,
  title={{USEV}: Universal speaker extraction with visual cue},
  author={Pan, Zexu and Ge, Meng and Li, Haizhou},
  journal=taslp,
  volume={30},
  pages={3032--3045},
  year={2022},
}

@inproceedings{lin2023avsepformer,
  title={{AV-SepFormer}: Cross-attention {SepFormer} for audio-visual target speaker extraction},
  author={Lin, Jiuxin and Cai, Xinyu and Dinkel, Heinrich and Chen, Jun and Yan, Zhiyong and Wang, Yongqing and Zhang, Junbo and Wu, Zhiyong and Wang, Yujun and Meng, Helen},
  booktitle=icassp,
  year={2023},
}

@inproceedings{sato2021multimodal,
  title={Multimodal attention fusion for target speaker extraction},
  author={Sato, Hiroshi and Ochiai, Tsubasa and Kinoshita, Keisuke and Delcroix, Marc and Nakatani, Tomohiro and Araki, Shoko},
  booktitle={IEEE Spoken Language Technology Workshop},
  year={2021},
}

@article{li2023rethinking,
  title={Rethinking the visual cues in audio-visual speaker extraction},
  author={Li, Junjie and Ge, Meng and Cao, Rui and Wang, Longbiao and Dang, Jianwu and Zhang, Shiliang and others},
  journal={arXiv preprint arXiv:2306.02625},
  year={2023}
}

@inproceedings{pan2023scenario,
  title={Scenario-aware audio-visual {TF-GridNet} for target speech extraction},
  author={Pan, Zexu and Wichern, Gordon and Masuyama, Yoshiki and Germain, Fran{\c{c}}ois G and Khurana, Sameer and Hori, Chiori and Le Roux, Jonathan},
  booktitle=asru,
  year={2023},
}

@inproceedings{gao2021visualvoice,
  title={{VisualVoice}: Audio-visual speech separation with cross-modal consistency},
  author={Gao, Ruohan and Grauman, Kristen},
  booktitle=cvpr,
  year={2021}
}

@article{li2024ctcnet,
  title={An audio-visual speech separation model inspired by cortico-thalamo-cortical circuits},
  author={Li, Kai and Xie, Fenghua and Chen, Hang and Yuan, Kexin and Hu, Xiaolin},
  journal=tpami,
  year={2024},
}

@article{luo2019conv,
  title={{Conv-TasNet}: Surpassing ideal time--frequency magnitude masking for speech separation},
  author={Luo, Yi and Mesgarani, Nima},
  journal=taslp,
  volume={27},
  number={8},
  pages={1256--1266},
  year={2019},
}

@article{li2024spmamba,
  title={{SPMamba}: State-space model is all you need in speech separation},
  author={Li, Kai and Chen, Guo and Yang, Runxuan and Hu, Xiaolin},
  journal={arXiv preprint arXiv:2404.02063},
  year={2024}
}

@inproceedings{luo2020dprnn,
  title={{Dual-path RNN}: Efficient long sequence modeling for time-domain single-channel speech separation},
  author={Luo, Yi and Chen, Zhuo and Yoshioka, Takuya},
  booktitle=icassp,
  year={2020},
}

@inproceedings{wang2023tf,
  title={{TF-GridNet}: Making time-frequency domain models great again for monaural speaker separation},
  author={Wang, Zhong-Qiu and Cornell, Samuele and Choi, Shukjae and Lee, Younglo and Kim, Byeong-Yeol and Watanabe, Shinji},
  booktitle=icassp,
  year={2023},
}

@inproceedings{zhao2024mossformer2,
  title={{MossFormer2}: Combining transformer and {RNN}-free recurrent network for enhanced time-domain monaural speech separation},
  author={Zhao, Shengkui and Ma, Yukun and Ni, Chongjia and Zhang, Chong and Wang, Hao and Nguyen, Trung Hieu and Zhou, Kun and Yip, Jia Qi and Ng, Dianwen and Ma, Bin},
  booktitle=icassp,
  year={2024},
}

@inproceedings{jiang2025dpmamba,
  title={{Dual-path Mamba}: Short and long-term bidirectional selective structured state space models for speech separation},
  author={Jiang, Xilin and Han, Cong and Mesgarani, Nima},
  booktitle=icassp,
  year={2025},
}

@article{cosentino2020librimix,
  title={{LibriMix}: An open-source dataset for generalizable speech separation},
  author={Cosentino, Joris and Pariente, Manuel and Cornell, Samuele and Deleforge, Antoine and Vincent, Emmanuel},
  journal={arXiv preprint arXiv:2005.11262},
  year={2020}
}

@inproceedings{chung2017lip,
  title={Lip reading sentences in the wild},
  author={Chung, Joon Son and Senior, Andrew W and Vinyals, Oriol and Zisserman, Andrew and others},
  booktitle=cvpr,
  year={2017}
}

@inproceedings{le2019sdr,
  title={{SDR}--half-baked or well done?},
  author={Le Roux, Jonathan and Wisdom, Scott and Erdogan, Hakan and Hershey, John R},
  booktitle=icassp,
  year={2019},
}

@inproceedings{reddy2021dnsmos,
  title={{DNSMOS}: A non-intrusive perceptual objective speech quality metric to evaluate noise suppressors},
  author={Reddy, Chandan KA and Gopal, Vishak and Cutler, Ross},
  booktitle=icassp,
  year={2021},
}

@inproceedings{mittag2021nisqa,
  title={{NISQA}: A deep {CNN}-self-attention model for multidimensional speech quality prediction with crowdsourced datasets},
  author={Mittag, Gabriel and Naderi, Babak and Chehadi, Assmaa and M{\"o}ller, Sebastian},
  booktitle=is,
  year={2021}
}

@inproceedings{kingma2015adam,
  author = {Kingma, Diederik P. and Ba, Jimmy},
  title = {Adam: A Method for Stochastic Optimization},
  booktitle = iclr,
  year = {2015},
}

@inproceedings{chung2018voxceleb2,
  title={{VoxCeleb2}: Deep speaker recognition},
  author={Chung, Joon Son and Nagrani, Arsha and Zisserman, Andrew},
  booktitle=is,
  year={2018}
}

@inproceedings{wu2019time,
  title={Time domain audio visual speech separation},
  author={Wu, Jian and Xu, Yong and Zhang, Shi-Xiong and Chen, Lian-Wu and Yu, Meng and Xie, Lei and Yu, Dong},
  booktitle=asru,
  year={2019},
}

@inproceedings{chou2024av2wav,
  title={{AV2Wav}: Diffusion-based re-synthesis from continuous self-supervised features for audio-visual speech enhancement},
  author={Chou, Ju-Chieh and Chien, Chung-Ming and Livescu, Karen},
  booktitle=icassp,
  year={2024},
}

@InProceedings{Chung16a,
  author       = "Chung, J.~S. and Zisserman, A.",
  title        = "Out of time: automated lip sync in the wild",
  booktitle    = "Workshop on Multi-view Lip-reading, ACCV",
  year         = "2016",
}

@inproceedings{prajwal2020lip,
  title={A lip sync expert is all you need for speech to lip generation in the wild},
  author={Prajwal, KR and Mukhopadhyay, Rudrabha and Namboodiri, Vinay P and Jawahar, CV},
  booktitle=acmmm,
  year={2020}
}

@article{yamagishi2019vctk,
  title={{CSTR VCTK Corpus}: English multi-speaker corpus for {CSTR} voice cloning toolkit (version 0.92)},
  author={Yamagishi, Junichi and Veaux, Christophe and MacDonald, Kirsten},
  journal={The Rainbow Passage which the speakers read out can be found in the International Dialects of English Archive},
  year={2019},
  publisher={University of Edinburgh. The Centre for Speech Technology Research (CSTR)}
}

@inproceedings{zen2019libritts,
  title={{LibriTTS}: A corpus derived from librispeech for text-to-speech},
  author={Zen, Heiga and Dang, Viet and Clark, Rob and Zhang, Yu and Weiss, Ron J and Jia, Ye and Chen, Zhifeng and Wu, Yonghui},
  booktitle=is,
  year={2019}
}

@inproceedings{yu2017permutation,
  title={Permutation invariant training of deep models for speaker-independent multi-talker speech separation},
  author={Yu, Dong and Kolb{\ae}k, Morten and Tan, Zheng-Hua and Jensen, Jesper},
  booktitle=icassp,
  year={2017},
}

\end{document}